\documentclass{iaus}
\usepackage{graphicx}

\title[MAGPHYS: a tool to interpret observed galaxy SEDs] %% give here short title %%
{MAGPHYS: a publicly available tool to interpret observed galaxy SEDs}

\author[E. da Cunha et al.]   %% give here short author list %%
{Elisabete da Cunha$^1$,
 St\'ephane Charlot$^2$,
Loretta Dunne$^3$,
Dan Smith$^{4}$,
 \and Kate Rowlands$^3$}

\affiliation{$^1$Max Planck Institute for Astronomy, K\"onigstuhl 17, 69117 Heidelberg, Germany \\ email: {\tt cunha@mpia.de} \\[\affilskip]
$^2$ UPMC Univ. Paris 6/CNRS, UMR 7095, Institut d'Astrophysique de Paris, France \\[\affilskip]
$^3$ School of Physics \& Astronomy, Nottingham University, University Park Campus, Nottingham NG7 2RD, UK\\[\affilskip]
$^4$ Centre for Astrophysics, Science \& Technology Research Institute, University of Hertfordshire, Hatfield, Herts, AL10 9AB, UK
}

\pubyear{2011} %% 
\volume{284}  %% insert here IAU Symposium No.
\pagerange{1--12}
\jname{The Spectral Energy Distribution of Galaxies}
\editors{R.J. Tuffs \&  C.C.Popescu, eds.}
\begin{document}

\maketitle

\begin{abstract}
We present a simple, physically-motivated model to interpret consistently the emission from galaxies at ultraviolet, optical and infrared wavelengths.
We combine this model with a Bayesian method to
obtain robust statistical constraints on key parameters describing the stellar content, star formation activity and dust content of galaxies.
Our model is now publicly available via a user-friendly code package, MAGPHYS at {\sf www.iap.fr/magphys}.
We present an application of this model to interpret a sample of $\sim1400$ local ($z<0.5$) galaxies from the H-ATLAS survey. We find that, for these galaxies,
the diffuse interstellar medium, powered mainly by stars older than 10~Myr, accounts for about half the total
infrared luminosity. We discuss the implications of this result to the use of star formation rate indicators based on total infrared luminosity.
\end{abstract}

\firstsection % if your document starts with a section,
\section{A simple model to interpret galaxy SEDs}

Multi-wavelength surveys of large samples of galaxies both in the local and high-redshift Universe have become 
widely available in recent years. To understand these observations in the framework of
galaxy formation and evolution, we must be able to extract key physical parameters
 from their observed spectral energy distributions (SEDs).
In \cite{daCunha2008}, we have developed a simple model to interpret consistently the ultraviolet,
optical and infrared emission from galaxies in terms of their star formation histories and dust content.

\subsection{Description of the model}

We compute the spectral evolution of stellar populations in galaxies using the state-of-the-art population synthesis model
of \cite{Bruzual2003}. This model is based on the property that stellar populations with any star formation history can be expanded
in a series of instantaneous bursts, Ôsimple stellar populationsÕ (SSPs). The spectral energy distribution of a galaxy is then
computed by adding the individual spectra of all SSPs weighted by the star formation rate over time since the galaxy was formed.

\begin{figure*}
\begin{center}
\includegraphics[width=\textwidth]{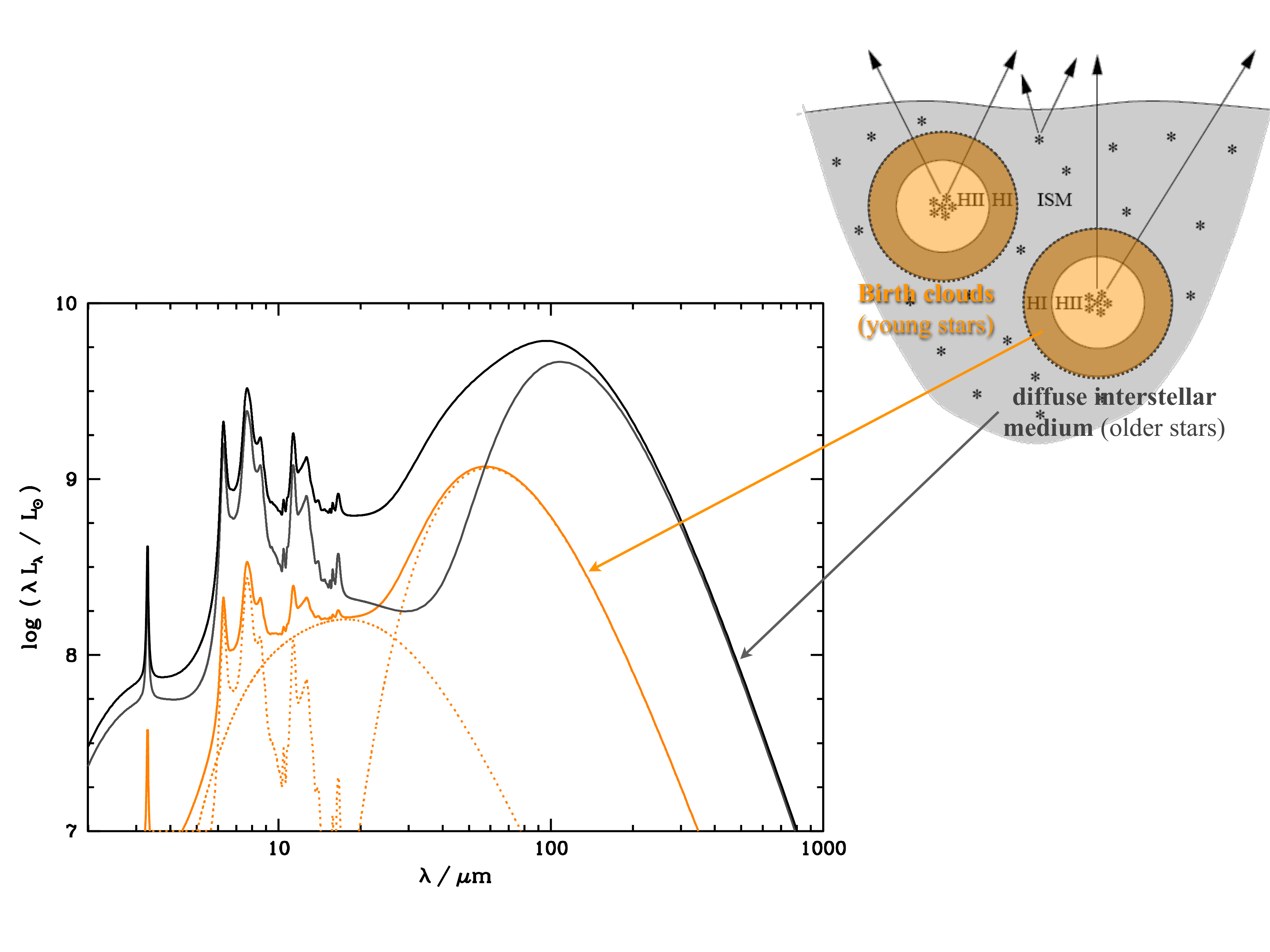}
\vspace{-1cm}
\caption{Schematic view of the two-component ISM by \cite{Charlot2000} (top-right): stars are born in dense molecular clouds -- birth clouds -- and later (after $10^7$~yr) migrate to
the ambient diffuse ISM. The left-bottom plot shows an example total dust emission SED (in black) in the infrared range, constructed using the emission components described in \cite{daCunha2008}.
The contributions by the birth clouds and the ambient ISM are plotted in orange and grey, respectively.}
\label{ism}
\end{center}
\end{figure*}

The spectra of galaxies also contain valuable information about the interaction of starlight
with interstellar gas and dust, and the physical properties of the interstellar medium (ISM), such as dust content.
Following \cite{Charlot2000}, we describe the ISM of galaxies in our model using two main components:
the ambient (diffuse) interstellar medium and the star-forming regions (birth clouds). Stars are born in dense molecular clouds
which dissipate typically on a time-scale of $10^7$ years. As a result, the non-ionizing continuum emission from young OB stars and
line emission from their surrounding H{\sc ii} regions may be absorbed by dust in these birth clouds and then in the ambient ISM, while the
light emitted by stars older than $10^7$~yr propagates only through the diffuse ISM. This simple model successfully accounts for the different
attenuation of line and continuum emission in star-forming galaxies.
We use this prescription to compute the total energy absorbed by dust in the birth clouds and in the ambient ISM.
We define the total dust luminosity re-radiated by dust in the birth clouds and in the ambient ISM as $L_\mathrm{d}^\mathrm{\,BC}$ and $L_\mathrm{d}^\mathrm{\,ISM}$, respectively. The total luminosity emitted by dust in the galaxy is $L_\mathrm{d}^\mathrm{\,tot}=L_\mathrm{d}^\mathrm{\,BC}+L_\mathrm{d}^\mathrm{\,ISM}$. We distribute $L_\mathrm{d}^\mathrm{\,BC}$ and $L_\mathrm{d}^\mathrm{\,ISM}$ in wavelength over the range from 3 to 1000 $\mu$m using four main dust components:
{\it (i)}  the emission from polycyclic aromatic hydrocarbons (PAHs);
{\it (ii)} the mid-IR continuum from hot dust;
{\it (iii)} the emission from warm dust (30--60~K) in thermal equilibrium;
{\it (iv)} the emission from cold dust (15--25~K) in thermal equilibrium.
In stellar birth clouds, the relative contributions to $L_\mathrm{d}^\mathrm{\,BC}$ by PAHs, the hot mid-infrared
continuum and warm dust are kept as adjustable parameters. These clouds are assumed not to contain any cold dust. In the ambient ISM, the contribution to $L_\mathrm{d}^\mathrm{\,ISM}$ by cold dust
is kept as an adjustable parameter. The relative proportions of the other 3 components are fixed to the values reproducing the mid-infrared cirrus emission of the Milky Way. We find that this minimum number of components is required to account for the infrared SEDs of galaxies in a wide range of star formation histories (see da Cunha et al. 2008 for details).

\subsection{Statistical constraints of physical parameters}

\begin{figure*}[b]
\begin{center}
\includegraphics[width=0.9\textwidth]{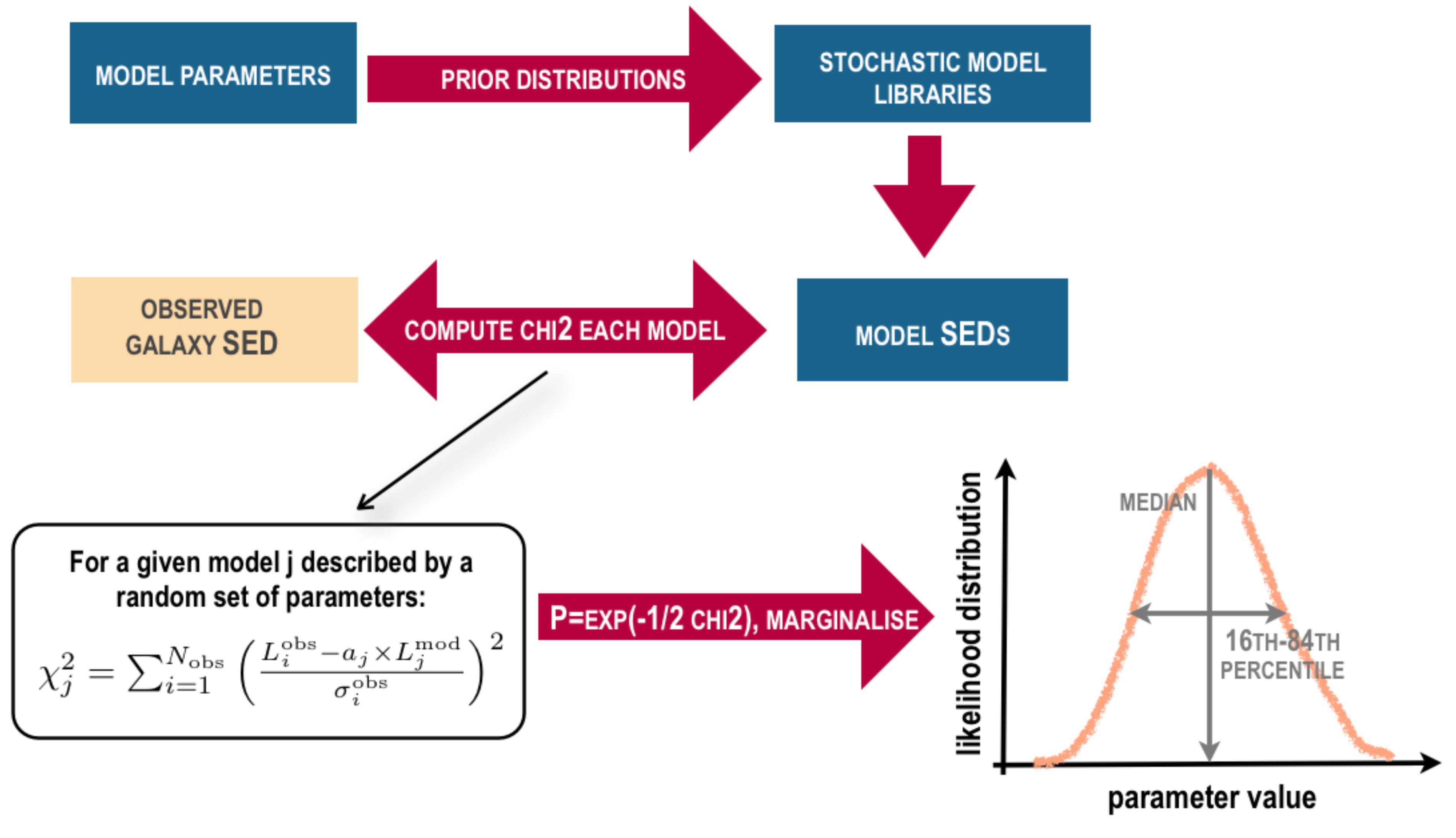}
\caption{Summary of the Bayesian approach we use to compare observed galaxy SEDs with our model and derive statistical constraints on the physical parameters of observed galaxies (see da Cunha et al. 2008 for details).}
\label{bayes}
\end{center}
\end{figure*}

Our model is optimized to derive statistical constraints of fundamental parameters related to star formation activity and dust content (e.g. star formation rate, stellar mass, dust attenuation, dust temperatures) of large samples of galaxies using a wide range of multi-wavelength observations (e.g. da Cunha et al. 2008, 2010).
We use a Bayesian approach, summarized in Fig.~\ref{bayes}, to interpret the SEDs all the way from the ultraviolet/optical  to the far-infrared. A similar approach has been previously used mostly to interpret optical galaxy spectra from the ultraviolet to the near-infrared, i.e. not including the dust emission.
This approach allows us to understand in detail our parameter constraints by identifying degeneracies and exploring what observations are required to constrain each parameter.

\section{The MAGPHYS package}

The \cite{daCunha2008} model is {\bf publicly available as the MAGPHYS package at {\sf www.iap.fr/magphys}}. MAGPHYS - {\em Multi-wavelength Analysis of Galaxy Physical Properties} - is a self-contained, user-friendly model package to interpret observed spectral energy distributions of galaxies in terms of galaxy-wide physical parameters pertaining to the stars and the interstellar medium.
The analysis of the spectral energy distribution of an observed galaxy with MAGPHYS is done in two steps: (i) The assembly of a comprehensive library of model SEDs at the same redshift and in the same photometric bands as the observed galaxy, for wide ranges of plausible physical parameters pertaining to the stars and ISM. (ii) The build-up of the marginalised likelihood distribution of each physical parameter of the observed galaxy, through the comparison of the observed SED with all the models in the library (Fig.~\ref{bayes}). 

\section{Dust heating by stars older than 10~Myr in H-ATLAS galaxies}

\begin{figure*}
\begin{center}
\includegraphics[width=0.65\textwidth]{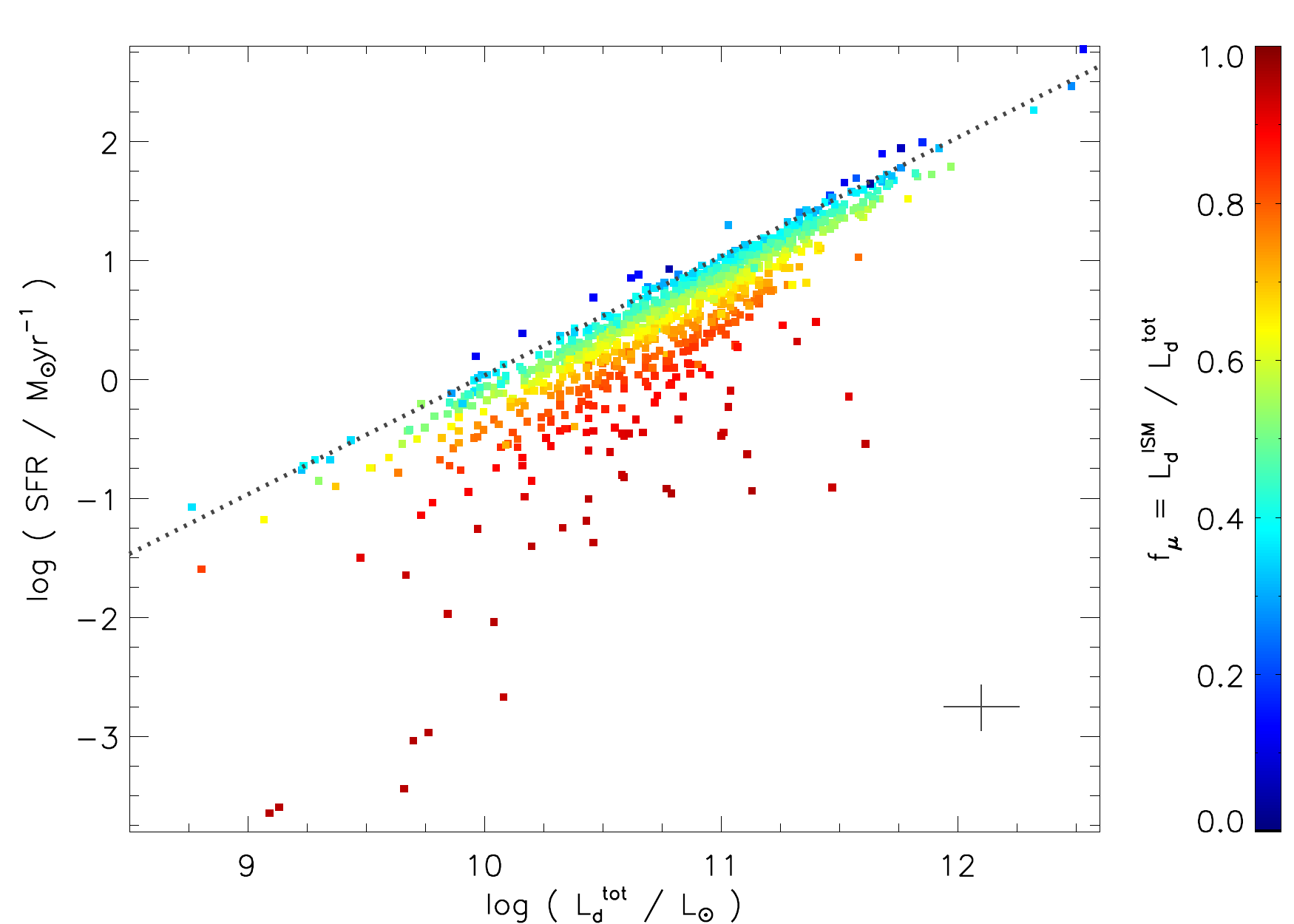}
\caption{Star formation rate, SFR, plotted against total dust luminosity, $L_\mathrm{d}^\mathrm{\,tot}$, for 1404 galaxies from the H-ATLAS survey (da Cunha et al., in prep.). Each galaxy is colour-coded according to the fraction of total infrared luminosity contributed by the diffuse ISM, $f_\mu$.
The dotted line shows, for reference, the conversion between total IR luminosity and SFR from \cite{Kennicutt1998}. }
\label{hatlas}
\end{center}
\end{figure*}

The H-ATLAS survey is detecting thousands of galaxies in the Herschel/SPIRE bands over a large area of the sky. Thanks to the overlap with other multi-wavelength surveys, complete UV-to-IR SEDs are available for large samples of galaxies (\cite{Eales2010}). We have used MAGPHYS to extract statistical constraints on star formation rates (SFRs), stellar masses, dust luminosities and dust masses of a sample of 250-$\mu$m detected galaxies at $z<0.5$ from their observed SEDs (Smith et al., subm.).
We are investigating the use of the IR luminosity as a SFR indicator in these galaxies.
The total IR is often used as a SFR tracer but observational evidence shows that dust in galaxies is not exclusively heated by newly-formed stars (e.g. \cite{Bendo2010}). We find that the SFR derived from fits to the total SED with our model, which includes heating by stars older than $10^7$~yr in the diffuse ISM, is not exactly traced by the dust luminosity (as, for example, when using the \cite{Kennicutt1998} IR to SFR conversion; see Fig.~\ref{hatlas}).
 A significant fraction (typically 50\%) of $L_\mathrm{d}^\mathrm{\,tot}$ in H-ATLAS galaxies comes from the diffuse ISM, mainly powered by stars older than 10~Myr. This implies that using the total IR as a SFR tracer may lead to overestimating the SFR unless the contribution by the diffuse ISM to the total IR is properly taken into account (see also e.g. \cite{Kennicutt2009}).

\acknowledgements
We thank the H-ATLAS team for providing the photometry for the work described in Section~3; the H-ATLAS website
is www.h-atlas.org.


\vspace{-0.5cm}
\begin{thebibliography}{}

\bibitem[Bendo et al. (2010)]{Bendo2010}
{Bendo} G. J., et al.,  2010, A\&A, 518, L65

\bibitem[Bruzual \& Charlot (2003)]{Bruzual2003}
{Bruzual} G.,  {Charlot} S.,  2003, MNRAS, 344, 1000

\bibitem[Charlot \& Fall (2000)]{Charlot2000}
{Charlot} S.,  {Fall} S.~M.,  2000, ApJ, 539, 718

\bibitem[da Cunha, Charlot \& Elbaz (2008)]{daCunha2008}
{da Cunha} E., {Charlot} S., {Elbaz} D.,  2008, MNRAS, 388, 1595

\bibitem[da Cunha et al. (2010)]{daCunha2010}
{da Cunha} E.,  {Eminian} C., {Charlot} S.,  {Blaizot} J.,  2010, MNRAS, 403, 1894

\bibitem[Eales et al. (2010)]{Eales2010}
{Eales} S., et al., 2010, PASP, 122, 499

\bibitem[Kennicutt (1998)]{Kennicutt1998}
{Kennicutt} R.C., Jr., 1998, ARA\&A, 36, 189

\bibitem[Kennicutt (2009)]{Kennicutt2009}
{Kennicutt} R.C., Jr., et al., 2009, ApJ, 703, 1672



\end{thebibliography}
\end{document}